\documentclass[fleqn,usenatbib]{mnras}

\usepackage{newtxtext,newtxmath}

\usepackage[T1]{fontenc}

\DeclareRobustCommand{\VAN}[3]{#2}
\let\VANthebibliography\thebibliography
\def\thebibliography{\DeclareRobustCommand{\VAN}[3]{##3}\VANthebibliography}

\usepackage{graphicx}	
\usepackage{amsmath}	

\newcommand{\ssax}{SAX~J1808.4$-$3658}
\newcommand{\sigr}{IGR~J17498$-$2921}

\newcommand{\nodata}{\ldots}

\title[System parameters for AMSP IGR J17498--2921]{Inferring system parameters from the bursts of the accretion-powered pulsar IGR J17498--2921}

\author[D. K. Galloway et al.]{
D. K. Galloway,$^{1,2,3}$\thanks{E-mail: duncan.galloway@monash.edu}
A. J. Goodwin,$^{4}$
T. Hilder,$^{1}$ 
L. Waterson$^{1}$
and 
M. Cup\'ak$^{5,6,4}$
\\
$^{1}$School of Physics and Astronomy, Monash University, Victoria 3800, Australia\\
$^{2}$OzGRav-Monash, School of Physics and Astronomy, Monash University, Victoria 3800, Australia\\
$^{3}$Institute for Globally Distributed Open Research and Education (IGDORE)\\
$^{4}$International Centre for Radio Astronomy Research – Curtin University, GPO Box U1987, Perth, WA 6845, Australia \\
$^{5}$Curtin Institute for Data Science, Curtin University, GPO Box U1987, Perth, WA 6845, Australia \\
$^{6}$Space Science and Technology Centre, School of Earth and Planetary Sciences, Curtin University, GPO Box U1987, Perth, WA, 6845, Australia
}

\date{Accepted 21/10/2024. Received 18/10/2024; in original form 22/3/2024}

\pubyear{2024}

\begin{document}
\label{firstpage}
\pagerange{\pageref{firstpage}--\pageref{lastpage}}
\maketitle

\begin{abstract}
Thermonuclear (type-I) bursts exhibit properties that depend both on the local  surface conditions of the neutron stars on which they ignite, as well as the physical parameters of the host binary system. However, constraining the system parameters requires a comprehensive method to compare the observed bursts to simulations. We have further developed the {\sc beansp} code for this purpose and analysed the bursts observed from IGR~J17498$-$2921, a 401-Hz accretion-powered pulsar, discovered during it's 2011 outburst. We find good agreement with a model having H-deficient fuel with $X=0.15\pm0.4$, and CNO metallicity 
$Z=0.0014^{+0.0004}_{-0.0003}$,
about a tenth of the solar value.
The model has the system at a distance of $5.7^{+0.6}_{-0.5}$~kpc, with a massive ($\approx2\ M_\odot$) neutron star and a likely inclination of $60^\circ$.
We also re-analysed the data from the 2002 outburst of the accretion-powered millisecond pulsar SAX~J1808.4$-$3658.
For that system we find a substantially closer distance than previously inferred, at $2.7\pm0.3$~kpc, likely driven by a larger degree of burst emission anisotropy. The other system parameters are largely consistent with the previous analysis.
We briefly discuss the implications for the evolution of these two systems.
\end{abstract}

\begin{keywords}
X-rays: binaries -- X-rays: bursts -- pulsars: individual: IGR~J17498$-$2921 -- pulsars: individual: SAX~J1808.4$-$3658 -- software: data analysis
\end{keywords}

\section{Introduction}

Accretion-powered millisecond pulsars (AMSPs) are a high observational priority due to their rarity \cite[only 20 are known;][]{disalvo22} and their significance as the evolutionary precursors of ``recycled'' millisecond radio pulsars \cite[e.g.][]{alpar82}. This link has been confirmed in spectacular fashion with the discovery in recent years of ``transitional'' pulsars which switch between rotation-powered (radio) and accretion-powered (X-ray) pulsar phases
{ \cite[]{papitto13b}.}
Millisecond radio pulsars (MSRPs) are thought to go through an AMSP stage, but present AMSPs have a distinctly different orbital period distribution compared to present-day MSRPs \cite[e.g.][]{tauris12,papitto14}.

AMSPs are typically discovered when they undergo a transient outburst, during which the accretion rate (and hence the X-ray luminosity) increases by several orders of magnitude in $\approx1$~d \cite[e.g.][]{goodwin20b}. This behaviour is also observed in the broader class of low-mass X-ray binaries, but the peak accretion rates reached by the AMSPs is typically lower.

Thermonuclear (type-I) bursts have been observed during the outbursts of some AMSPs, and are of particular interest due to their utility for constraining the properties of the host systems \cite[e.g.][]{gal21a}. The bursts from AMSPs typically exhibit burst oscillations at the pulse (spin) frequency \cite[]{pw12}. They tend to be of short duration, indicative of H-deficient fuel, and with a high incidence of photospheric radius-expansion. These properties are generally associated with accretion at low (few \% Eddington) rates, as inferred from the persistent X-ray emission.

\ssax\ was the first AMSP discovered \cite[]{zand98c,wij98b,chak98d}, and has been observed in outburst a total of 11 times \cite[e.g.][]{illiano23}.
The bursts from \ssax\ are perhaps the best-studied of all. First discovered with the {\it BeppoSAX} { Wide-Field Camera }\cite[WFC;][]{zand98c}, subsequent events have been observed by several missions, including simultaneously with {\it Chandra}\/ \& { the {\it Rossi X-ray Timing Explorer }} \cite[{\it RXTE};][]{zand13a} demonstrating the boost in persistent flux associated with bursts \cite[]{worpel13a,worpel15}. With the { Neutron-star Interior Composition Explorer} (NICER) a complex 2-stage flux evolution in the rise has been observed, implying compositional interactions with the radius-expansion phenomenon (\citealt{bult19b}; see also \citealt{gal06a}). The 2019 outburst, which was detected unusually early thanks to coordinated optical and X-ray monitoring \cite[]{goodwin20b}, revealed with NICER and { the {\it Nuclear Spectroscopic Telescope Array}} ({\it NuSTAR}) unusually weak bursts, attributed to H-ignition rather than the usual He \cite[]{casten23}.

Several attempts have been made to match the bursts from \ssax\ to 1-D ignition models, beginning with \cite{gal06c}. \cite{johnston18} generated plausible sequences of bursts for the same outburst with the {\sc kepler} code. \citet[][hereafter G19]{goodwin19c} developed a Bayesian framework ({\sc beansp}) and constrained the fuel composition, as well as the system parameters, demonstrating that the mass donor was H-deficient. These  constraints also have consequences for the evolution of the system \cite[]{goodwin20a}.

In order to further develop the capabilities of the {\sc beansp} code, in this work we applied it to the 2011 outburst of \sigr. 
This pulsar towards the Galactic centre ($l=0.16$, $b=-1.00$) was discovered by the {\it INTErnational Gamma-Ray Astrophysics Laboratory}\/ ({\it INTEGRAL}) when it went into outburst in 2011 August \cite[]{papitto11b}, and was active for $\approx40$~d subsequently. The source consists of a neutron star rotating with a spin frequency of 401~Hz, and a binary companion orbiting once every 3.8~hr. Thermonuclear bursts were detected during the outburst, first by the Joint European X-Ray Monitor (JEM-X) instrument onboard {\it INTEGRAL} \cite[]{ferrigno11}, as well as subsequently with the {\it RXTE} Proportional Counter Array (PCA) and {\it Swift}. The bursts were short duration (6--10~s) and were consistent with recurrence times in the range 16--18~hr. As with other AMSPs, the bursts exhibited burst oscillations at a frequency consistent with the persistent pulsations,  with only one having evidence for photospheric radius expansion. The implied distance (neglecting any burst anisotropy) is in the range 6.6--8.1~kpc.
The source was detected once again in outburst in 2023 April \cite[]{grebenev23,sanna23} 
{ with observations by {\it NuSTAR}, {\it NICER}, {\it Insight-HMXT}\/ and {\it INTEGRAL}\/ indicating renewed bursting activity \cite[]{li24,bhatt24a,illiano24b}. }

We reiterate here the motivation for choosing AMSPs as the target for these analyses, { and also the 
{ specific burst ignition  code for comparison with the observations}}. First, of the broader population of low-mass X-ray binaries, typically more information is known about AMSPs, specifically their orbital and rotation periods. 
{ Second, the varying accretion rate during the outburst provides the opportunity to sample bursts at different accretion rates, which might serve to resolve degeneracies in the model-observation comparisons.
Third,} the relatively low ($<5$\% of Eddington) maximum accretion rates achieved during their outbursts favour so-called ``class III'' bursts \cite[He ignition in a H-depleted fuel layer;][]{gal21a}. { Because stable (hot CNO) H-burning no longer contributes to the temperature around the ignition depth,} the ignition of such events is less dependent on the burning from previous bursts; at higher accretion rates the ``class V'' (mixed H/He) bursts typical of the ``clocked'' burster, GS~1826$-$24 \cite[]{gal03d} depend significantly on the { compositional and thermal ``inertia'' (i.e., the nuclear burning history) following the previous event.}

{ We chose the {\sc settle} code for burst ignition calculations primarily for pragmatic purposes, as it is much faster to run than fully time-dependent simulations. It was originally used to predict the hydrostatic expansion of the burning shell during bursts \cite[]{cb00}. The code integrates the temperature profile from the surface inward through the burning layer, with simple energy generation prescription that depends primarily on fuel composition (H-fraction $X$ and CNO metallicity $Z$) and the accretion rate, $\dot{m}$. The integration terminates at a depth where the conditions for unstable ignition are met; an additional ``base flux'' term ($Q_b$) accounts for possible heat arising from deeper layers.
Thus, as {\sc settle} does not simulate the burning between the bursts, we expect (as also argued by \citealt{gal06c}) that it will reproduce the ``class III'' bursts typical for AMSPs with relatively higher fidelity than other classes. }

In this paper we describe our efforts to constrain the system parameters for \sigr\ via comparison of the observed bursts with a numerical ignition model. In \S\ref{sec:methods} we describe the comparison approach, via the {\sc beansp} code.
In \S\ref{sec:obs} we describe the observational data used for the analysis, and give more detail about how the Markov-Chain Monte Carlo (MCMC) algorithm is used to constrain the system parameters.
In \S\ref{sec:results} we describe the results of the MCMC runs, and the corresponding constraints derived on the target systems.
Finally, in \S\ref{sec:discussion} we discuss the results and the next steps.

\section{Methods}
\label{sec:methods}

We adopted the same approach as G19 for matching bursts to numerical models, but have made substantial improvements to the {\sc beansp}\footnote{\url{https://github.com/adellej/beans}} code base,
{as described further in  appendix \ref{sec:improvements}.

{The {\sc beansp} code generates sequences of bursts based on the inferred accretion rate, calculated in turn from the observed persistent flux history of the target source. 
{ Each burst is triggered when the conditions for unstable helium burning are met at the base of the accumulated fuel layer, at a temperature of $\approx2\times10^8$~K and a density range of $~10^5 -10^6\ {\rm g\,cm^{-3}}$. }
The model input parameters determining the properties of the bursts are 
\begin{equation}
(X, Z, Q_b, d, \xi_b, \xi_p) \label{eq:inputs}
\end{equation}
i.e. the fuel H-fraction $X$ and CNO metallicity $Z$; the ``base'' flux $Q_b$, in units of MeV~nucleon$^{-1}$; the distance $d$, in kpc; the burst and persistent emission anisotropy, $\xi_b$ and $\xi_p$, defined
}
in the same sense as for \cite{he16}, i.e. the flux measured by an observer $F_{b,p}$ is related to the total luminosity $L_{b,p}$ by
\begin{equation}
F_{b,p} = \frac{L_{b,p}}{4\pi d^2 \xi_{b,p}}
\end{equation}
For example, $\xi_{b,p}>1$ indicates that the emission is beamed preferentially out of the observer's line of sight, and the observed flux is lower than would be expected for isotropic emission.
{ The accretion rate per unit area $\dot{m}$ is derived from the persitent flux $F_p$ via 
\begin{equation}
    \dot{m} = F_p c_{\rm bol} \frac{d^2\xi_p(1+z)}{R^2Q_{\rm grav}} 
\end{equation}
where $c_{\rm bol}$ is the bolometric correction, and $Q_{\rm grav}=c^2z/(1+z)$ is the gravitational energy release per gram, with $z$ the  gravitational redshift at the neutron star surface \cite[cf. with][ equation 6]{concord22}.}
{The neutron star mass and radius are fixed by default at $M=1.4\ M_\odot$ and $R=11.2$~km, respectively; the user may optionally also allow these parameters to vary.} 

{ For each burst interval, the code uses an iterative method to predict the recurrence time, necessary because {\sc settle} requires a constant input accretion rate $\dot{m}$, but the accretion rate for the AMSP targets varies with time. Beginning at some burst time $t_i$, an initial  estimate of $\dot{m}'_0$ determines, with the other input parameters (equation \ref{eq:inputs}), a trial burst interval $\Delta t'_0$. We then average the persistent flux over the interval $(t_i, t_i+\Delta t'_0)$ to provide an updated accretion rate $\dot{m}'_1$ and (via another {\sc settle} run) a new recurrence time estimate $\Delta t'_1$. We repeat this process until there is no further change to $\Delta t'_i$, i.e. a self-consistent solution is provided where the average inferred accretion rate $\dot{m}_{i+1}$ over the burst interval corresponds precisely with the predicted burst interval $\Delta t_{i+1}$ \cite[see][for a possible caveat for this approach]{johnston18}. The code carries out this procedure both forwards and backwards from a ``reference'' burst to cover the entire extent of the burst observations.}

{\sc beansp} uses the {\sc emcee} implementation of the affine-invariant Markov-Chain Monte-Carlo (MCMC) sampler \cite[]{emcee13} { to draw and evolve samples from the parameter posterior distributions}.
{ The {\sc beansp} code itself }
is now available via the Python Package Index\footnote{\url{http://pypi.org}} (PYPI).
The burst matching algorithm relies on the ignition code of \cite{cb00}, which has now been improved and updated and is available as {\sc pySettle}\footnote{\url{https://github.com/adellej/pysettle}} also  via PYPI. 
}

\section{Observations \& Simulations}
\label{sec:obs}

As a primary source we used burst analyses from the Multi-INstrument Burst ARchive \cite[MINBAR;][]{minbar}, which includes analyses of all bursts observed with the {\it RXTE}\/ Proportional Counter Array (PCA), {\it BeppoSAX}/WFC and the {\it INTEGRAL}\/ Joint European Monitor for X-rays (JEM-X) through to 2012 January. Where necessary, we augmented these data with additional burst and persistent flux measurements observed with instruments not contributing to that sample.  

The burst measurements { include} the start time, typically measured as the time that the burst flux first exceeded 25\% of the maximum reached during the burst. The burst fluence $E_b$ is calculated from time-resolved spectroscopy covering the burst, 
{ integrating a decay model into the tail where required to ensure complete coverage.}
The $\alpha$-value, where used, is calculated as
\begin{equation}
    \alpha_i = \frac{F_{p,i} c_{\rm bol}\Delta t_i}{E_{b,i}}
\end{equation}
where $F_{p,i}$ is the average persistent flux over the interval from the previous burst, $E_{b,i}$ is the fluence of burst $i$, and $c_{\rm bol}$ is the bolometric correction required for the flux, which is usually measured in a restricted energy band (3--25~keV for the instruments contributing to MINBAR).
{ The burst recurrence time $\Delta t_i$ is a key parameter for determining $\alpha_i$, but it may not always be possible to determine unambiguously from observations alone. The largest uncertainties arise due to the possibility of bursts occurring during data gaps; such missed bursts may also bias the persistent flux measurement over the interval (see also \S\ref{sec:sim}).}

In order to verify the new version of the code against the old, we re-analysed the bursts from the 2002 outburst of \ssax\ as well as the 2011 outburst of \sigr, as described below.

\subsection{SAX J1808.4--3658}

We adopted the same data used by G19, obtained from {\it RXTE}/PCA observations of the 2002 outburst and analysed as part of the MINBAR sample.
The data comprises four bursts and persistent flux measurements with {\it RXTE}/PCA covering the interval 2002 October 15–-23 (MJD 52562--52570; Table \ref{tab:s1808_bursts}). 
As the persistent fluxes are measured from spectral fits in the 3--25~keV range, we adopted a constant bolometric correction of 2.21 over the course of the outburst, { following G19}.
We used the spline approximation to integrate the persistent flux between bursts for the purposes of burst simulation, with smoothing factor 0.02. 
{ This value was chosen to reduce the possible impact of local maxima in the flux history, which may prevent convergence of the iterative scheme to determine the burst trains (cf. with \S\ref{sec:methods}).}

\subsection{IGR J17498--2921}

\begin{figure}
	\includegraphics[width=\columnwidth]{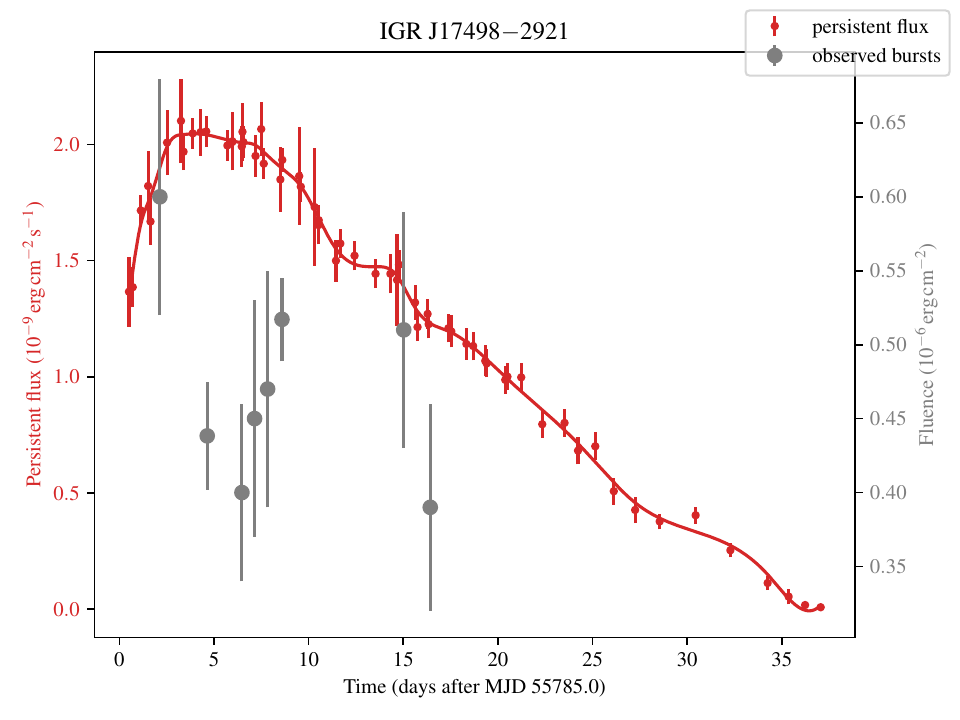}
    \caption{Observations of \sigr\ by {\it RXTE}/PCA, {\it INTEGRAL}/ISGRI and {\it Swift}/XRT during the 2011 outburst. The persistent flux in the 0.1--300~keV band estimated from spectral fits by \protect\cite{falanga12} are shown as red points, and the observed burst fluences are shown as grey. Error bars indicate the $1\sigma$ uncertainties.
    Note the much longer duration of the outburst, and the larger number of detected bursts, compared to \ssax.
    The spline approximation ({\tt smooth=0.1}) used to integrate the persistent flux between bursts for the purposes of burst simulation is indicated as the solid red line.
    \label{fig:igr_outburst} }
\end{figure}

We adopted the  persistent fluxes in the 0.1--300~keV range estimated by \cite{falanga12}, covering the outburst interval of 2011 August 12--September 21 (MJD~55785--55825; Fig. \ref{fig:igr_outburst}). As the energy band is already sufficiently wide we take those fluxes as bolometric, and so do not require an additional correction. Those authors also reported eight bursts, two observed each with {\it RXTE}/PCA and {\it Swift}, and five with {\it INTEGRAL}/JEM-X (one of those was also observed with {\it Swift}).

As the persistent flux measurements were somewhat noisy, we chose the ``spline'' option for integrating the flux between bursts, with the smoothing factor of 0.1 chosen to strike a balance between preserving the overall features of the lightcurve, but not introducing additional (possibly spurious) variations from outlier measurements. To achieve this we also omitted three flux measurements, two by {\it Swift} and one by {\it INTEGRAL} following the peak of the outburst.

\subsection{Simulations}
\label{sec:sim}

The simulation approach involves three principal steps. First, identifying a suitable parameter vector as a start point. The primary input parameters to the model are the hydrogen fraction $X$ and CNO metallicity $Z$; the base flux $Q_b$; the distance $d$ and anisotropy factors $\xi_b$ and $\xi_p$; and the neutron-star mass $M_{\rm NS}$ and radius $R_{\rm NS}$. The ``systematic'' error multipliers on the $\alpha$ and fluence values are two additional optional parameters, that are not used in the runs presented here (except where specified). 

A qualitatively good initial solution will approximately replicate the burst history, including the burst times and number. Because the {\sc beansp} code works both forward and backward in time from a ``reference'' burst, some adjustment of the reference burst choice ({\tt ref\_ind}) and the number of bursts to simulate ({\tt numburstssim}) may also be required.
For \ssax, this exercise is trivial, as we can use the derived parameters from G19, along with the same parameters.

Second, running the MCMC sampler and identifying promising solution regions. Typically, running the sampler with 200 walkers for up to 1000 steps will provide a balance between run time and exploration of the parameter space. As the burst rate varies for different walker positions, the total number of bursts simulated at each step may also vary. Promising solution regions can be identified by plotting the burst predictions (which are divided up automatically as part of the {\tt do\_analysis} method based on the burst number) and comparing the RMS error between the average predicted burst times and the observations.

A key feature of this simulation approach is the necessity to match the predicted bursts with those observed. Because the data coverage is usually incomplete, there is a high probability of some bursts being missed; e.g. the pair of bursts inferred to have occurred between the first and second observed bursts in the 2002 outburst of \ssax\ (G19, \citealt{gal06c}). To compare the observed burst properties with the predictions, a burst in the predicted sequence must be selected for each observed event. The {\sc beansp} code achieves this by selecting the mapping that minimises the RMS offset between the observed times and the predicted times for the selected bursts. However, as the walker positions evolve during the simulation, the mapping will sometimes change from step to step, likely introducing discontinuities in the likelihood hypersurface. Practically the consequences are that the confidence regions may be strongly multi-modal, and caution must be taken to ensure that the globally best-fit solution has been achieved.

Third, refining the walker ensemble to focus on the preferred solution. The user can ``prune'' the set of walkers after a preliminary run to remove those with sub-optimal burst matching sequences, and continue the simulation to fully explore the parameter space around the best identified solution.

The {\sc beansp} code can take advantage of multi-core CPUs by using the {\tt threads} option to run in parallel for as many threads as are available. The walker positions are adjusted each step with the {\sc emcee} default ``stretch move'' of \cite{gw10}, with default scale parameter $a=2.0$. For many of the runs presented here, the sampler gave an error which we understand is related to an insufficiently low fraction of successful moves (the average acceptance fraction at the end of the ``base'' runs for each source was 7--8\%). We overcame this issue by successively reducing the scale factor in steps as $a'=a/2+1/2$ (i.e. 1.5, 1.25, 1.125 etc). Typically the runs were completed with $a=1.25$.

{ The runs presented in this paper were performed by comparing only the burst start times and fluences, neglecting the measured $\alpha$-values.
The $\alpha$-values are typically calculated from the persistent fluxes 
measured since the earlier burst; because the instrumental coverage may not be complete, this quantity 
may be an under- or over-estimate of the averaged flux over that interval. Furthermore, for some events the inferred recurrence time may be subject to uncertainties from the number of missed bursts since the previous event. 
We allowed the neutron star mass and radius to vary, but omitted the multiplicative factors $f_i$, included by G19 to account for possible underestimation of the variance in the fluence. }

\section{Results}
\label{sec:results}

Here we describe the results from our simulations. In each case, the run parameters and posterior samples are available as accompanying data with the paper, at the companion repository site (see 
the ``data availability'' section).

\subsection{SAX J1808.4--3658}
\label{sec:s1808}

We repeated the analysis of G19,
running the MCMC sampler with 500 walkers and (initially) 2000 steps.
We omitted the $\alpha$ values in the likelihood, for the reasons described in \S\ref{sec:methods} (see also below). We also omitted 
the multiplicative factor for the fluence uncertainties.

We could not replicate the calculated autocorrelation times from G19; for our runs the initial values cluster around $\tau=10$ initially, rising linearly seemingly independent of the length of the run, and not reaching the target criteria even well beyond the 2000 steps of the previous runs.
For that reason, we continued the MCMC chains for 8000 additional steps, for a total of 10000, and retained the last 1000 steps for the posteriors. The estimated $\tau\approx1000$ by the end of this run, so still not meeting the rule-of-thumb for convergence. 

\begin{figure}
	\includegraphics[width=\columnwidth]{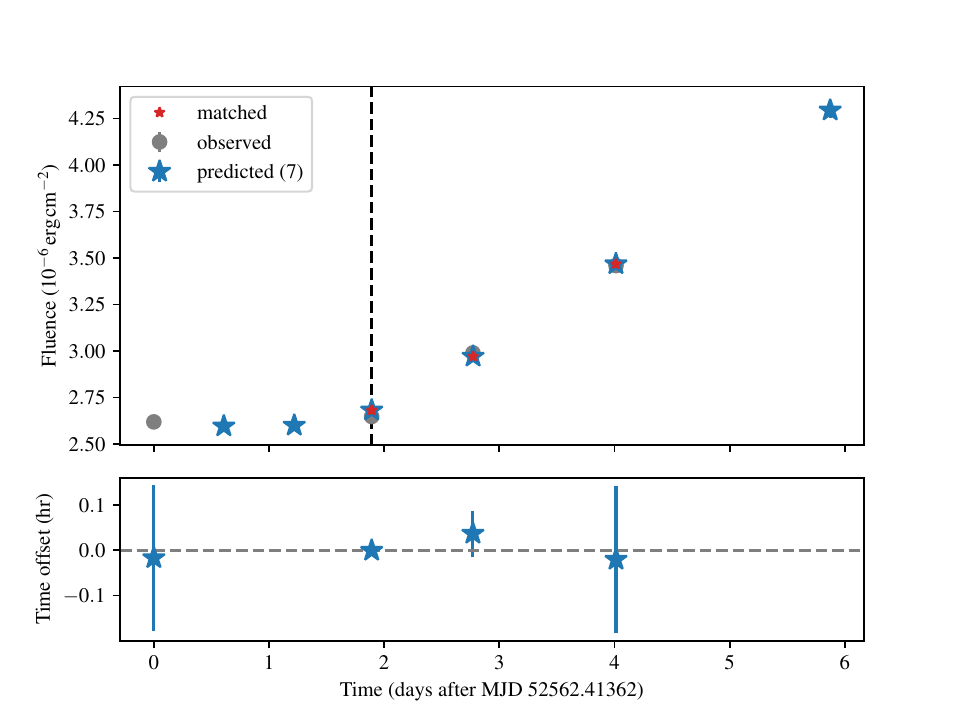}
    \caption{Comparison of observed and 
    { model predicted} burst times and fluences for \ssax\ during the 2002 outburst. The top panel plots fluence as a function of burst times; the predicted bursts ({\it blue stars}) chosen for comparison with the observed bursts ({\it grey circles}) are additionally marked by a red star.
    Note the unobserved bursts between the first two observed events, and following the fourth. These events would have fallen during data gaps.
    The ``reference'' burst (from which the simulation proceeds both forward and backward in time) is marked as the black dashed line
    The lower panel shows the residuals between the observed and predicted times. There is no error on the second observed burst, as it is the reference for the simulations and hence is trivially identical to the observed time.
    \label{fig:sax_bursts} }
\end{figure}

The average of the predicted burst times and fluences match the observations well, with a residual RMS of 1.37~min (Fig. \ref{fig:sax_bursts}). 
This agreement is somewhat remarkable given the adopted error on the burst time of 10~min; in fact from the PCA data the start time of the bursts can be constrained to well under 1~s.
As with previous analyses, we predict the occurrence of two additional bursts between the first two observed bursts; the predicted times fell within intervals during which no observations were available, so are still consistent with the observations. We estimated the recurrence time for the second observed burst from the difference of it's observed time and the time of the third predicted burst (Table \ref{tab:s1808_bursts}).

\begin{figure*}
	\includegraphics[width=0.7\textwidth]{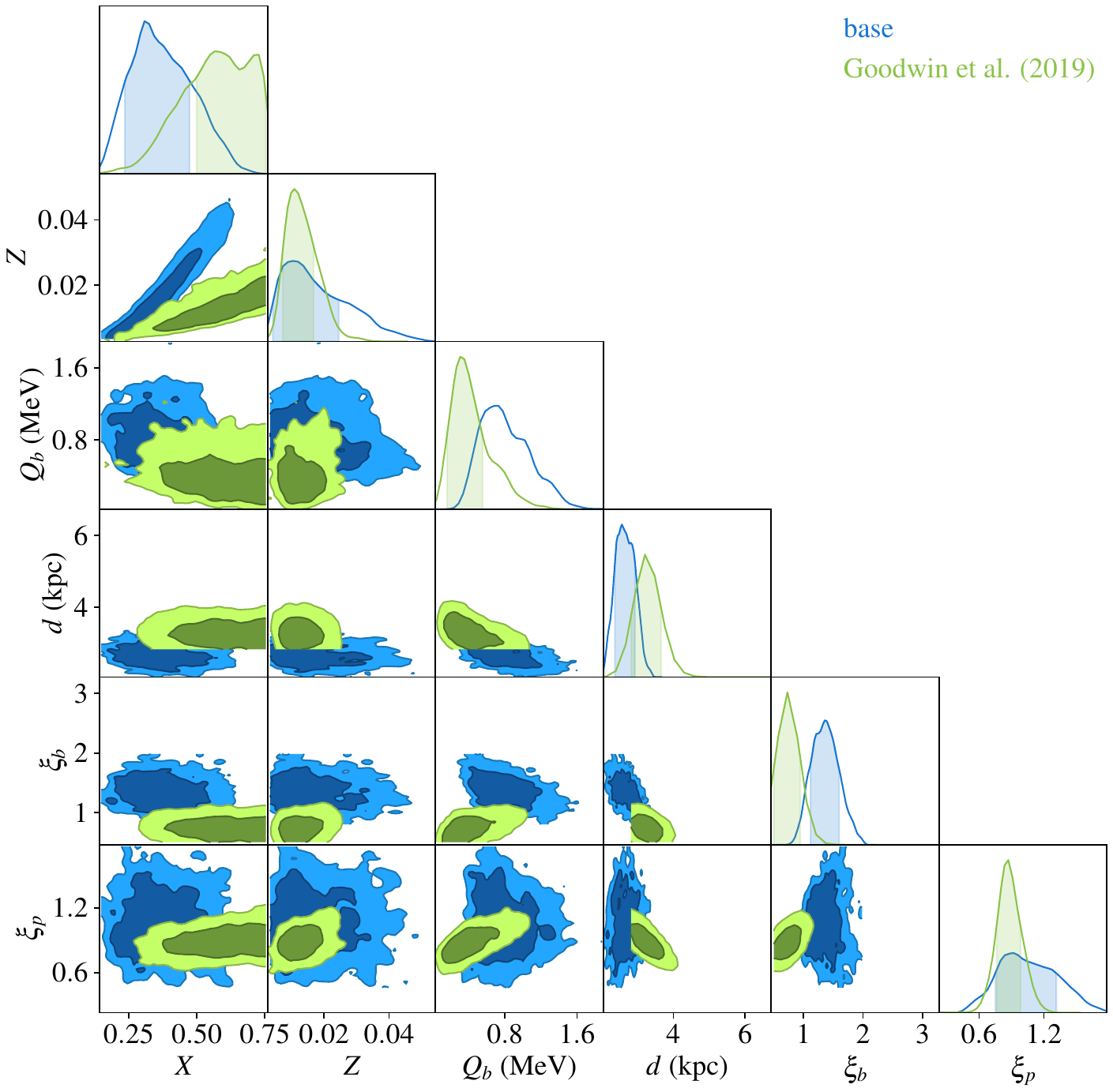}
    \caption{Two-dimensional posterior probabilities for the system parameters for \ssax. The blue distributions show the results for our ``base'' run as described in \S\ref{sec:s1808}, while the green distributions are those of G19.
    The displayed confidence contours within each 2-D panel are at 1 and $2\sigma$.
    The displayed parameters are the hydrogen mass fraction $X$, metallicity $Z$, base flux $Q_{\rm b}$, distance $d$, and the emission anisotropy parameters $\xi_b$ and $\xi_p$.
    %
    \label{fig:s1808_pos} }
\end{figure*}

The calculated posteriors are broadly consistent with those of the previous analysis (Fig. \ref{fig:s1808_pos}). We find again the correlation between $X$ and $Z$ first noted by \cite{gal06c}, 
{ arising from the mutual dependence of hot-CNO burning on these parameters, to achieve the required H-fraction at ignition }.
Interestingly, the track in the $X$--$Z$ plane has shifted to higher $Z$ values { compared to the previous analysis}. We attribute this shift to the changes made to the code and the comparison algorithm, as described in \S\ref{sec:methods}.

Comparing the tabulated values, we note the posterior ranges have substantial overlap with the previous ranges (Table \ref{tab:s1808_params}). We find that the hydrogen fraction $X$ is now much better constrained, with the value of $X=0.36$ having the maximum 
{ posterior probability.}
{The detailed structure measured in the rise of other radius-expansion bursts from this source suggests a higher H-fraction, of $\approx0.7$ \cite[]{bult19b,guichandut23}; however, our posterior distribution is quite broad and does not exclude this value at high significance. }

The mass posterior is shifted to much higher values, centred around $2.1\ M_\odot$; along with the slightly higher radius, these parameters lead to higher values of the surface gravity $g$ and redshift $1+z$, although not significantly discrepant from the previous analysis. The implied distance is somewhat closer, at $2.7\pm0.3$~kpc.
We also note a higher value for the burst anisotropy $\xi_b$, which could be the principal reason for the correspondingly lower distance.
{The photospheric radius-expansion observed for these bursts may indicate that a  fraction of the burst energy is unobservable, lost as kinetic energy in the temporarily expanded atmosphere \cite[e.g.][]{yu18}. The fraction of energy lost will contribute to the $\xi_b$ value, provided that it is roughly constant from burst to burst.}  

We revisit here the choice to omit the $\alpha$-values from the model-observation comparison. 
Also in Table \ref{tab:s1808_bursts} we compare the measured $\alpha$-values with the calculation incorporating both the model-predicted recurrence time (since the previous burst, whether observed or predicted) and the variations in the persistent flux over the entire interval (according to our flux model). We calculated the $\alpha$-values and propagated the uncertainties using routines from the {\sc concord} suite of tools \cite[]{concord22}. We note that the measured values substantially underestimate the model-informed values, which further supports our choice (at least in this case) to omit them from the likelihood calculation.

\begin{table*}
	\centering
	\caption{Measured and inferred properties for bursts from \ssax\/ 
 { observed with {\it RXTE}/PCA} 
 during the 2002 outburst, augmented by the model predictions. We infer the presence of two additional bursts between the first two observed events; hence, the recurrence time $\Delta t$ for the second event depends on the predictions, and thus introduces uncertainties. Similarly, the $\alpha$-values include this uncertainty, as well as the variation in persistent flux according to the model.}
	\label{tab:s1808_bursts}
	\begin{tabular}{ccccccc} 
		\hline
        & MINBAR & Start  & $\Delta t$ & Fluence & \multicolumn{2}{c}{$\alpha$-value} \\
  Burst & ID     & (MJD)  & (hr)       & $10^{-6}\ {\rm erg\,cm^{-2}}$ & Measured & Inferred \\
		\hline 
1 & 3037 & 52562.41362 & \nodata & $2.620 \pm 0.021$ & \nodata & \nodata \\
2 & 3038 & 52564.30514 & $16.13\pm0.05$ & $2.649 \pm 0.018$ & $107.0 \pm 2.0$ & $117.5\pm0.9$\\
3 & 3039 & 52565.18427 & 21.10 & $2.990 \pm 0.017$ & $105.7 \pm 1.9$ & $121.7_{-0.6}^{+0.8}$\\
4 & 3040 & 52566.42677 & 29.82 & $3.460 \pm 0.022$ & $121.1 \pm 2.3$ & $130.2\pm0.8$\\
		\hline
    \end{tabular}
\end{table*}

\begin{table}
	\centering
	\caption{System parameters for \ssax\ and \sigr\ derived from the burst comparison.}
	\label{tab:s1808_params}
	\begin{tabular}{cccl} 
		\hline
  Parameter & \ssax\ & \sigr\ & Units \\
		\hline 
$X$ & $0.36^{+0.13}_{-0.11}$      & $0.15\pm0.04$ & \\
$Z$ & $0.016^{+0.014}_{-0.008}$   & $0.0014^{+0.0004}_{-0.0003}$ & \\
$Q_{\rm b}$ & $0.8^{+0.3}_{-0.2}$ & $2.1^{+0.6}_{-0.5}$ & MeV nucleon$^{-1}$ \\
$M$ &   $2.1^{+0.3}_{-0.4}$       & $2.2^{+0.2}_{-0.4}$ & $M_\odot$\\
$R$ & $12.2\pm0.9$                & $12.0^{+0.9}_{-1.1}$ & km\\
$g$ & $2.5\pm0.5$                 & $3.0^{+0.6}_{-0.5}$ & $10^{14}\ {\rm cm\,s^{-2}}$\\
$1+z$ & $1.41^{+0.08}_{-0.10}$    & $1.48^{+0.07}_{-0.09}$ & \\
$d$ & $2.7\pm0.3$                 & $5.7^{+0.6}_{-0.5}$ & kpc \\
$\xi_p$ & $1.0^{+0.3}_{-0.2}$     & $1.4\pm0.20$ & \\
$\xi_b$ & $1.4\pm0.2$             & $1.5^{+0.3}_{-0.2}$ & \\
		\hline
    \end{tabular}
\end{table}

\subsection{IGR J17498--2921}

The larger number of bursts for this source presented a much greater challenge for simulation. \cite{falanga12} suggested low inclination to explain the high $\alpha$ values, and we used this information (via the $\xi_b$ and $\xi_p$ parameters), along with the suggested distance range, to guide our initial choice of parameters. 

The walkers in the early simulations tended to fragment into separate groups corresponding to different burst rates and consequently different mappings from the predicted to observed bursts. We gradually refined the available choices, prioritising the agreement on the burst time, for the final runs contributing to the parameter estimates here.
For the final posteriors and model prediction ranges we chose a simulation running for 12,000 steps total, with 500 walkers. After 10,000 steps we ``pruned'' the set of walkers to retain only those which offered the best agreement in terms of the predicted times. The sub-optimal solutions, comprising only 5.7\% of the walkers, had RMS time offsets of $>2$~hr, compared to $<1$~hr for the remainder. We replaced each of the walkers with sub-optimal solutions, with one chosen randomly from the set exhibiting better agreement, and ran for an additional 2,000 steps. For the final quantities and plots we discarded the initial 11,000 steps as burnin.

As before, the integrated autocorrelation time estimates did not indicate that the run had yet converged, as for \ssax. The lack of convergence may be exacerbated by the low acceptance fraction for the walkers (see \S\ref{sec:sim}).

\begin{figure}
	\includegraphics[width=\columnwidth]{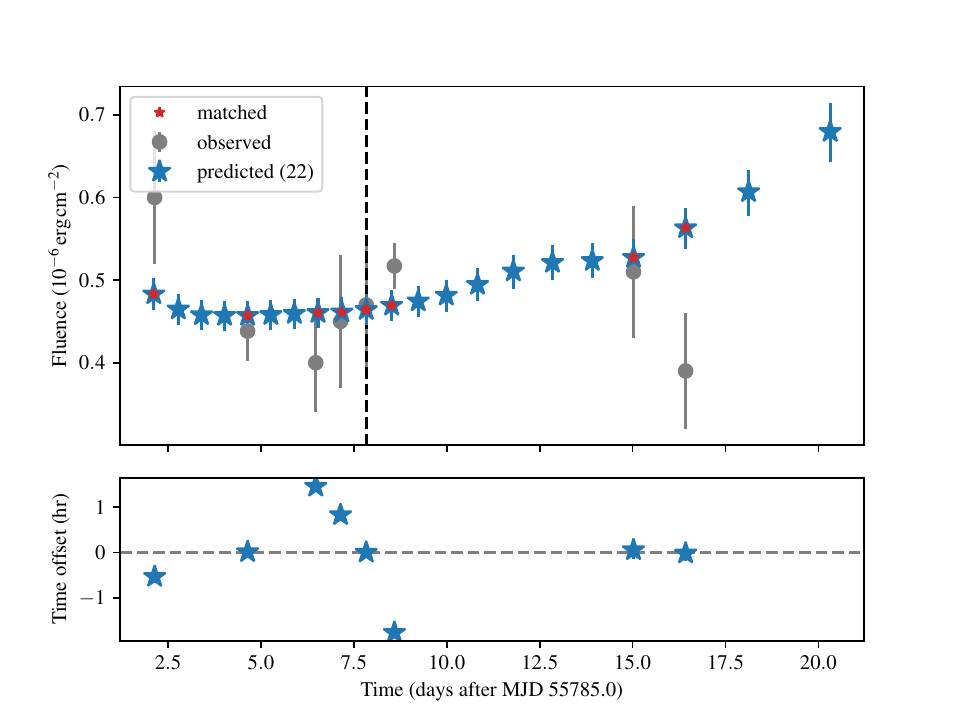}
    \caption{Comparison of observed and MCMC ensemble burst times and fluences for \sigr\ during the 2011 outburst. For this dataset the reference burst is chosen as the fifth observed event. The other details are as for Fig. 
    \ref{fig:sax_bursts} \label{fig:igr_bursts}}
\end{figure}

The comparison of predicted and observed times and fluences was good, with RMS 0.882~hr for the times (Fig. \ref{fig:igr_bursts}). However, the agreement was far from that of the \ssax\ bursts, and systematic variations apparently dominate the residuals. The observed variations in the fluences for the bursts from \sigr\ were also reproduced moderately well, generally consistent on average given the observational uncertainties. The predicted fluence was anticorrelated with the persistent flux (i.e. the inferred accretion rate).

The inferred burst recurrence times and $\alpha$-values are listed in Table \ref{tab:i17498_bursts}.
The burst recurrence times were also anticorrelated with the accretion rate, and varied between 13.85~hr at the peak of the outburst (between bursts 2 and 3) to 33.65~hr between bursts 7 and 8.
The consistency of the $\alpha$ values around 250 for the main part of the burst train is perhaps a good indicator of reliability. As pointed out by \cite{falanga12}, this value is well in excess of the maximum value expected for pure He bursts, of 150 \cite[see also][]{concord22}. However, the observed value  also includes the ratio $\xi_b/\xi_p$, which according to the posteriors is $\approx1.18$, bringing it closer to agreement.

It is interesting that the final observed burst in the train is at lower fluence than it's predecessor, the opposite of the trend for the predicted bursts. In response the $\alpha$-value is even higher, around 400, and significantly in excess of the values earlier in the outburst. Although systematic effects and incorrect instrumental cross-calibration may affect the fluence measurement of this burst (observed with {\it Swift}), if the fluence is reliable it may indicate either that the burning is incomplete, or also that steady burning is removing a higher fraction of the accreted fuel than earlier in the outburst.

\begin{figure}
	\includegraphics[width=\columnwidth]{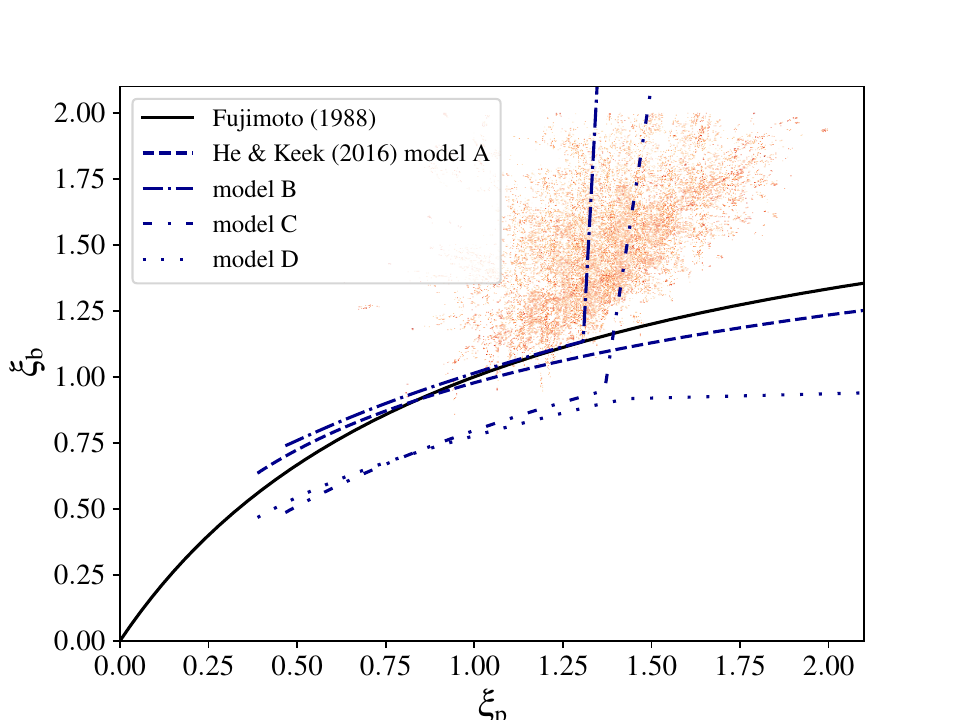}
    \caption{The inferred distribution for \sigr\ of burst and persistent emission anisotropy, $\xi_b$ and $\xi_p$ respectively, plotted against the predictions from analytic and numerical models. 
    Note that the $\xi_b$ vlaue is generally in excess of the range spanned for thin or flared disks, instead suggesting a trapezoidal (B) or triangular (C) cross-sectional profile.
    \label{fig:igr_xis}}
\end{figure}

The system parameters appear to largely support the case for high inclination, with both $\xi_b$ and $\xi_p$ well in excess of 1 (Fig. \ref{fig:igr_xis}). 
{With most of these bursts exhibiting no evidence of photospheric radius-expansion, we expect the sole contribution to $\xi_b$ to be the burst emission anisotropy (cf. with \S\ref{sec:s1808}). }
Interestingly, the preferred value of $\xi_b$ is in excess of typical ranges for the standard ``model A'', and instead falls in the region traced by the flared trapezoidal (B) or triangular (C) model profiles of \cite{he16}. 
The implied inclination is $60^\circ$ or higher, where the flared disk edge begins to obstruct the line of sight to the neutron star, substantially attenuating both the persistent and burst flux.
{ This value is consistent with the inferred range derived from persistent spectral modelling with a reflection component by \cite{illiano24b}, but not with the smaller values (30--40$^\circ$) inferred by \cite{bhatt24a} and \cite{li24} based on similar approaches. }

\begin{figure*}
	\includegraphics[width=0.7\textwidth]{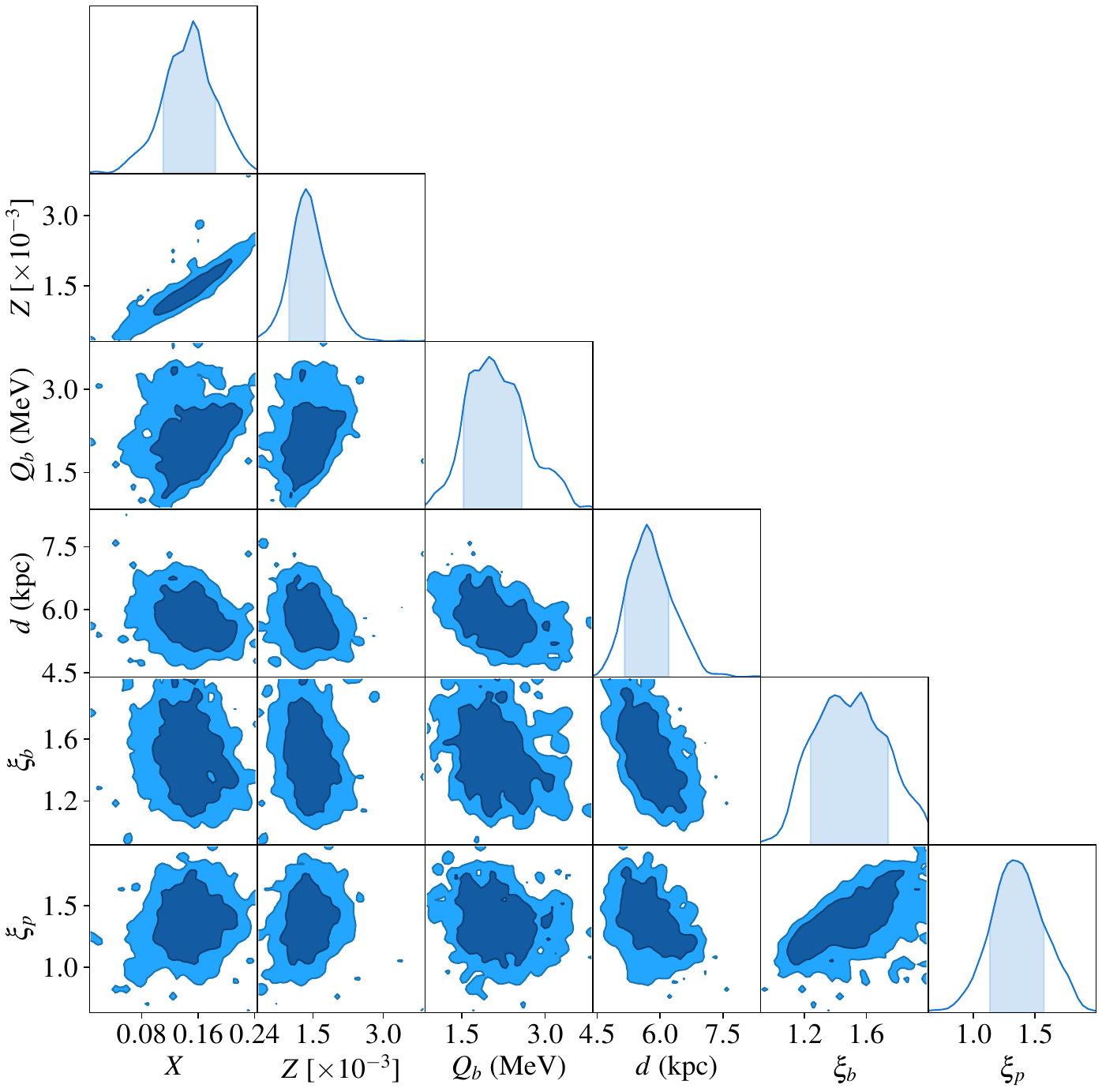}
    \caption{Two-dimensional posterior probabilities for the system parameters for \sigr. Details as for Fig. \ref{fig:s1808_pos} 
    \label{fig:igr_pos}}
\end{figure*}

The implied composition for the burst fuel is both H- and CNO-poor (Fig. \ref{fig:igr_pos}), with most remarkably the preferred $Z$ values an order of magnitude below solar. 
Perhaps to compensate, the implied base flux is higher than usually assumed, around 2~MeV~nucleon$^{-1}$.

The distance posterior is centred around 5.7~kpc, which is somewhat closer than the range considered by \cite{falanga12}. However, those authors did not take into account the emission anisotropy, which provides a significant correction given our best estimates. For that reason, we consider the closer distance more likely.

\begin{figure*}
	\includegraphics[width=0.7\textwidth]{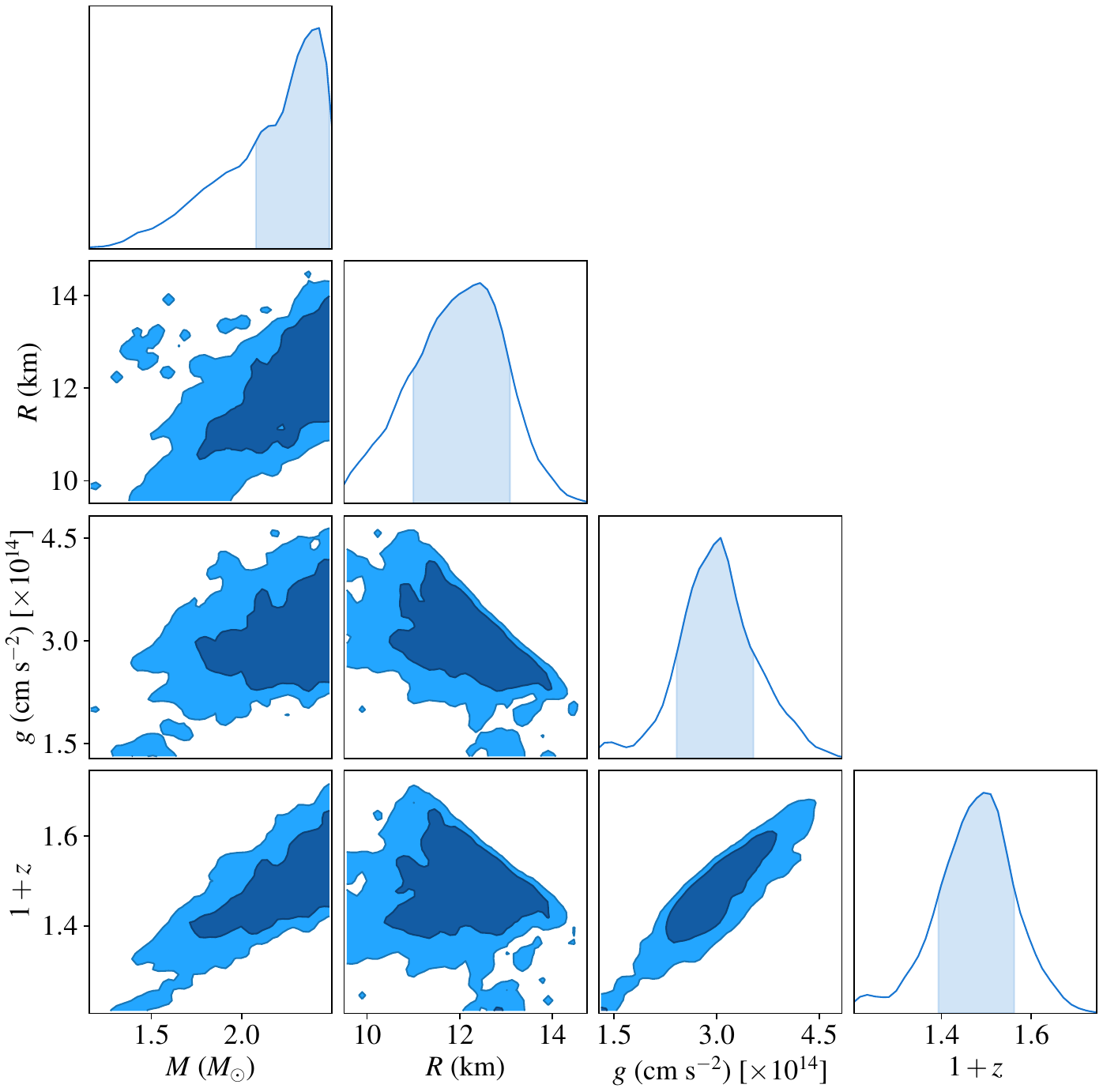}
    \caption{Two-dimensional posterior probabilities for the neutron star parameters for \sigr.  
    \label{fig:igr_mr}}
\end{figure*}

While the neutron-star radius is consistent with the canonical $\approx12$~km value, the mass distribution is strongly skewed towards high values (Fig. \ref{fig:igr_mr}).
When combined with our inclination estimate and the pulsar mass function $f (M_2, M_1, i) \simeq 2\times10^{-3}\ M_\odot$ \cite[]{papitto11b}, the implied donor is lower mass, around $0.2\ M_\odot$, and must have significantly larger radius than a zero-age main sequence star of the same mass, in order to fill it's Roche lobe.

\begin{table*}
	\centering
	\caption{Measured and inferred properties for bursts from \sigr\/ 
 { observed with {\it RXTE}/PCA, {\it INTEGRAL}/JEM-X and {\it Swift}\/ } during the 2011 outburst, augmented by the model predictions. { Burst numbers are as for those in Table 2 of \protect\cite{falanga12}. }
}
	\label{tab:i17498_bursts}
	\begin{tabular}{ccccccc} 
		\hline
        & MINBAR & Start  & $\Delta t$ & Fluence & \multicolumn{2}{c}{$\alpha$-value} \\
  Burst & ID     & (MJD)  & (hr)       & $10^{-6}\ {\rm erg\,cm^{-2}}$ & Measured & Inferred \\
		\hline 
1 & 8486 & 55787.13924 & $18.92\pm0.18$ & $0.60\pm0.08$ & $160\pm70$ & $200_{-20}^{+30}$ \\
2 & 8239 & 55789.64013 & $14.88\pm0.09$ & $0.44\pm0.04$ & $310\pm40$ & $250_{-19}^{+23}$ \\
3 & 8498 & 55791.47419 & $13.85\pm0.04$ & $0.40\pm0.06$ & $320\pm50$ & $250_{-30}^{+40}$ \\
4 & 8503 & 55792.14131 & 16.01 & $0.45\pm0.08$ & $290\pm60$ & $260_{-40}^{+50}$ \\
5 & 8508 & 55792.83207 & 16.58 & $0.47\pm0.08$ & $270\pm50$ & $250_{-40}^{+50}$ \\
6 & 8244 & 55793.5906 & 18.20 & $0.52\pm0.03$ & $250\pm20$ & $244_{-13}^{+14}$ \\
7 & 8526 & 55800.02405 & $26.59\pm0.16$ & $0.51\pm0.08$ & $230\pm40$ & $270_{-40}^{+50}$ \\
8 & \nodata & 55801.42595 & 33.65 & $0.39\pm0.07$ & $250\pm50$ & $400_{-60}^{+90}$ \\

		\hline
    \end{tabular}
\end{table*}

\section{Discussion}
\label{sec:discussion}

We have made a detailed comparison of the predictions of an ignition model to 8  bursts observed from \sigr\ during it's 2011 outburst, using the {\sc beansp} code. We can broadly reproduce the behaviour with a model in which the accretion fuel is H-poor, with $X=0.15\pm0.04$, and also low metallicity, at about a tenth of the solar value. The implied distance to the source is $5.7^{+0.6}_{-0.5}$~kpc, and the system is likely observed at an inclination of $\approx60^\circ$, with indications that a ``flared'' accretion disk substantially attenuates both the burst and persistent luminosity towards our line of sight.
The model predictions favour a massive neutron star, with $M\geq2\ M_\odot$, and $R\approx12$~km.

We achieved these constraints using an upgraded version of the {\sc beansp} code, featuring substantial upgrades and improvements over the earlier versions. To quantify the effect of these changes, we have also re-analysed the observations of \ssax\ analysed by G19. 
We find that in general the new code produces comparable parameter confidence intervals, although we could not reproduce the autocorrelation times. Thus, the convergence of the MCMC chains remains uncertain.

{ We note that convergence issues are commonly found in practical Baysesian inference \cite[e.g.][]{hfm18}, and can arise from a number of circumstances, including complex posterior geometries (including multiple local regions of high likelihood); lack of model flexibility, such that the model cannot closely replicate the data (perhaps due to additional systematic contributions to the variance); and underestimated uncertainties, which can also contribute to the preceding two effects. In particular for {\sc beansp} the discrete variations that may arise in the burst matching procedure (i.e. where the number of missed bursts between a pair of bursts may jump up or down by 1 due to variations in the model parameters) are likely to make convergence more difficult to achieve; other techniques such as nested sampling may be better suited to this application. }

We performed our comparisons on only the burst times and fluences, neglecting the estimated $\alpha$-values, as has been done in the past. By combining the predicted burst times with the observed fluences, we demonstrated that the $\alpha$ values estimated from the observations can have significant systematic errors, due to the varying persistent flux and uncertainties for the burst recurrence times.

We expect that the improvements in the code demonstrated here with the application to \sigr\ will enable further comparisons to burst trains from other sources.
{ It is hoped that additional applications along with further testing will help to resolve the issues with convergence and to fully quantify the sensitivity of the comparison as a function of sample size. }

\subsection{Implications for the binary evolution of SAX J1808.4--3658 and IGR~J17498--2921}

In this work, we find that the accreted fuel composition in both binaries is hydrogen depleted. Given that the secondary star in both systems is low mass (0.05\,M$_{\rm{\odot}}$ and $>$0.16\,M$_{\rm{\odot}}$ respectively, \citealt{chak98d,markwardt11}) this finding has important implications about the evolution of the binary systems. The present-day companions must represent the core of  previously more massive stars. In order to deplete hydrogen in the core within the age of the Universe, both companion stars must have been significantly more massive than they are today, implying that they have lost a significant amount of mass via accretion to the neutron star in their lifetimes. 

The binary evolution of \ssax\ was modelled by \citet{goodwin20a} taking into account the depleted hydrogen fraction of the companion star. Using the Modules for Experiments in Stellar
Astrophysics stellar evolution binary program 
\citep[MESA;][]{mesa15}
to calculate evolutionary tracks of the binary system, they found that the companion star likely had an initial mass of 1.1\,M$_{\rm{\odot}}$, but at least $>0.6$\,M$_{\rm{\odot}}$. 

Similarly for \sigr, using MESA to evolve a single star for 14\,billion\,yr, we deduce that the companion star must have been $>0.8$\,M$_{\rm{\odot}}$ in order to achieve a hydrogen fraction of 0.14 in its core within the age of the Universe. 
We plan more detailed simulation efforts in future to fully elucidate the evolutionary history of \sigr\/ as well as \ssax.

\section*{Acknowledgements}

This work was supported by software support resources awarded under the Astronomy Data and Computing Services (ADACS) Merit Allocation Program. ADACS is funded from the Astronomy National Collaborative Research Infrastructure Strategy (NCRIS) allocation provided by the Australian Government and managed by Astronomy Australia Limited (AAL).
Parts of this research were conducted by the Australian Research Council Centre of Excellence for Gravitational Wave Discovery (OzGrav), through project number CE170100004. 
This work was supported in part by the National Science Foundation under Grant No. PHY-1430152 (JINA Center for the Evolution of the Elements).
This work was supported by the Australian government through the Australian Research
Council’s Discovery Projects funding scheme (DP200102471).
This research has made use of data obtained through the High Energy Astrophysics Science Archive Research Center Online Service, provided by the NASA/Goddard Space Flight Center.

\section*{Data Availability}
\label{data}

The data underlying this article are available in Monash University's Bridges repository, at \url{https://dx.doi.org/10.26180/24773367}

\bibliographystyle{mnras}
\bibliography{all}

\appendix

\section{Improvements to the {\sc beansp} code}
\label{sec:improvements}

The burst matching algorithm of {\sc beansp} relies on the ignition code of \cite{cb00}, which has now been improved and updated and is available as {\sc pySettle}\footnote{\url{https://github.com/adellej/pysettle}} also  via PYPI. 
This code now gives correct values for the $\alpha$-parameter, i.e. the ratio of the persistent luminosity to the burst luminosity, in the observer frame for consistency with the burst recurrence time and energy.

The 0.65 correction factor adopted by G19 for the burst recurrence time $\Delta t$ and fluence $E_b$ has now been removed from the {\sc pySettle} code,
and incorporated instead as a correction function within {\sc beansp}. 
Examination of the code suggests that previously the factor was  applied only to the recurrence time, and the fluence was unchanged.
We provide the function {\tt corr\_goodwin19} to replicate the runs from this and the previous paper, but the user can also provide their own.

{\sc beansp} uses the {\sc emcee} implementation of the affine-invariant Markov-Chain Monte-Carlo (MCMC) sampler \cite[]{emcee13}.
Earlier versions of the code relied on conversion of model-predicted burst parameters to observational equivalents, via three ratios from which the physical parameters could be determined. 
Since the original version of the code used a fixed mass and radius, and the option to vary those parameters was introduced later, we cannot be confident that the conversion of the ratios to physical parameters correctly incorporated varying mass and radius.
The present version supplants this approach and replaces the three ratios with explicit physical parameters: distance $d$, and the emission anisotropies of the burst and persistent flux, $\xi_b$ and $\xi_p$, respectively. 

Other modifications to the code used in this work include corrections to the conversion between flux and accretion rate (parameterised in units of the Eddington rate, $1.75\times10^{-9}[1.7/(1+X)]\ M_\odot\,{\rm yr^{-1}}$ where $X$ is the hydrogen mass fraction) for {\sc pySettle}.
The user can now choose different options for integrating the persistent flux between bursts (to calculate the average accretion rate for the purposes of predicting the burst properties). Previously the code would interpolate linearly between flux measurements; this approach causes problems for the simulation algorithm due to local maxima (which may or may not be real) in the flux history. The user can instead choose to approximate the flux history with a B-spline representation \cite[{\tt splrep} in {\tt scipy};][]{scipy20}, with user-defined smoothing factor. 

The flexibility of the {\sc beansp} code has been substantially improved, facilitating exploration of the parameter space. The user can opt to omit the comparison of the measured $\alpha$ values, and only compare the burst times and fluences. This approach might be preferred for cases where the precise burst count through observing gaps cannot be maintained, so that the precise recurrence time is unknown. Additionally, there is the question of how independent the $\alpha$-values are, since they are calculated from the fluences combined with the recurrence time and the persistent fluxes. 
The user can further also omit the fluences from the comparison if an initial match purely on the burst times is desired, 
{or if the fluences cannot be measured (e.g. bursts observed with instruments with poor sensitivity or spectral resolution)}.
Bursts for which the fluence cannot be measured can still be included in the likelihood calculation on the basis of their start time alone.
Previously the likelihood calculation included multiplicative factors $f_i$ (see  equation 11 {of G19}) to account for possible underestimation of the variance. The present version can be run with or without these additional factors.

\bsp	
\label{lastpage}
\end{document}